# Experimental and numerical investigation on folding stable state of bistable deployable composite boom

Tian-Wei Liu[1,2,3], Jiang-Bo Bai[1,2]*, Hao-Tian Xi[1], Nicholas Fantuzzi[3], Guang-Yu Bu[1], Yan Shi[4]**

1. School of Transportation Science and Engineering, Beihang University, Beijing, 100191, People's Republic of China (*, corresponding author: baijiangbo@buaa.edu.cn)

2. Jingdezhen Research Institute of Beihang University, Jiangxi Province, 333000, People's Republic of China

3. DICAM Department, University of Bologna, Bologna 40136, Italy

4. School of Automation Science and Electrical Engineering, Beihang University, Beijing, 100191, People's Republic of China (**, corresponding author: shiyan@buaa.edu.cn)

**Abstract:** The bistable deployable composite boom (Bi-DCB) can realize the bistable function by storing and releasing strain energy, which has a good application prospect in the aerospace field. In this paper, the folding stable state of the Bi-DCB was investigated using experimental and numerical approaches. Using the vacuum bag method, six Bi-DCB specimens were prepared. Bistable experiments of Bi-DCB specimens were conducted and linear fitting with Archimedes' helix was performed to determine the folding stable configuration. In addition, two Finite Element Models (FEMs) were established for predicting the folding stable state of the Bi-DCB. Two classical failure criteria were utilized to analyze the stress level of the folding stable state of the Bi-DCB. Numerical results of two FEMs agreed well with experimental results, including the bistable deformation process and the folding stable state.

# Nomenclature

| | |
|---|---|
| $a$ | controls the distance between two adjacent circles |
| $b$ | distance from the start-point to the origin of the polar coordinate system |
| $D_{11}$ | bending stiffness of the laminate |
| $D_{12}$ | bending stiffness of the laminate |
| $d$ | diameter of the folding stable state predicted in reference, mm |
| $E_1$ | longitudinal elastic modulus, GPa |
| $E_2$ | transverse elastic modulus, GPa |
| $G_{12}$ | in-plane shear modulus, GPa |
| $G_{13}$ | inter-laminar shear modulus, GPa |
| $G_{23}$ | inter-laminar shear modulus, GPa |
| $I_{f,1}$ | Tsai-Hill failure index |
| $I_{f,2}$ | maximum stress failure index |
| $l$ | applied displacement, mm |
| $L$ | length of the Bi-DCB, mm |
| $L^*$ | displacement of rigid plate moving in $x$ direction, mm |
| $R$ | radius of cross-section, mm |
| $r_0$ | polar radius at the start-point of the Bi-DCB in the folding stable state, mm |
| $r_1$ | polar radius at the end-point of the Bi-DCB in the folding stable state, mm |
| $S_{12}$ | in-plane shear strength of composite ply, MPa |
| $X_t$ | longitudinal tensile strength of composite ply, MPa |
| $X_c$ | longitudinal compressive strength of composite ply, MPa |
| $X_1$ | intermediate variable, MPa |
| $X_2$ | intermediate variable, MPa |
| $Y_t$ | transverse tensile strength of composite ply, MPa |
| $Y_c$ | transverse compressive strength of composite ply, MPa |
| $Y$ | intermediate variable, MPa |
| $\alpha_{e,i}$ | polar angle corresponding to the $i^{\text{th}}$ data point on the neutral surface of the cross-section, rad |
| $\alpha$ | polar angle of fitted Archimedes' helix, rad |
| $\overline{\alpha_e}$ | average polar angle of 30 data points, rad |
| $\alpha_1$ | polar angle at the end-point of the fitted Archimedes' helix, rad |
| $\theta$ | central angle of the cross-section, deg |
| $\rho$ | polar radius of fitted Archimedes' helix, mm |
| $\rho'$ | density of composite, g/cm³ |
| $\rho_{e,i}$ | polar radius corresponding to the $i^{th}$ data point on the neutral surface of the cross-section, mm |
| $\overline{\rho_e}$ | average polar radius of 30 data points, |
| $\nu_{12}$ | Poisson's ratio |
| $\nu_{21}$ | Poisson's ratio |
| $\sigma_1^k$ | maximum longitudinal principal stress of the $k^{th}$ ply in the laminate, MPa |
| $\sigma_2^k$ | maximum transverse principal stress of the $k^{th}$ ply in the laminate, MPa |
| $\tau_{12}^k$ | maximum principal shear stress of the $k^{th}$ ply in the laminate, MPa |
| Bi-DCB | bistable deployable composite boom |
| CFRP | carbon-fiber-reinforced-plastics |
| DCB | deployable composite boom |
| FEM | finite element model |
| STEM | storable tubular extendible member |

## 1. Introduction

The deployable membrane structure has attracted increasing attention due to the advantages such as high storage ratio, light weight and excellent mechanical properties, and has successfully achieved engineering application or in orbit verification in large space deployable structures such as solar arrays, solar sails, parabolic antennas and inflatable cabins [1-3]. The deployable membrane structure is usually composed of the thin membrane and the deployable boom, which has the advantages of lightweight, low mass-to-volume ratio and high packaging efficiency. As the most important component of the deployable membrane structure, the deployable boom plays a supporting and tensioning role in the working state. The deployable boom is important to ensure smooth deployment and normal operation of the deployable membrane structure [4] (shown in Fig. 1).

Many early studies on the tubular deployable boom focused on isotropic materials, for example, Klein designed the thin-walled storable tubular extendible member (STEM) for the communication antenna of Canada's first satellite, the Alouette [5,6]. With the development of carbon-fiber-reinforced-plastics (CFRP) technology [7-14], the deployable composite boom (DCB) made of CFRP has obtained widespread attention. Stabile and Laurenzi [15] used nonlinear explicit dynamics analysis to simulate the folding behavior of the DCB. In the following work, the dynamic behavior of a DCB specially designed for the satellite platform was investigated through experiments and finite element method, and the experiments are in good agreement with numerical results [16]. Yang et al. [17] performed multi-objective optimization design for the DCB through the finite element method and response surface method. Fernandez et al. [18] designed and manufactured a derail sail with an area of 25 square meters, which could be folded and deployed by four DCBs. It is noting that the above deployable composite extension arms are all monostable structures, which are stable only in the deploying state. A complex folding mechanism is required for folding storage, which significantly increases the weight of spacecrafts and launch costs.

Through reasonable antisymmetric layup design, the DCB can have a bistable function, that is, there are two stable states in the deploying and folding configurations (shown in Fig. 1b). The Bi-DCB can maintain the folding stable state without any constraint. Moreover, the Bi-DCB has a smaller strain energy and shows a more controllable self-deployment function in the folding stable state. Therefore, the Bi-DCB has a wider application prospect. There are analytical method, numerical simulation and experiments to investigate the folding stable state of the Bi-DCB. For the analytic method, Iqbal et al. [19] first established a simple linear elastic bistable analytical model to predict the radius of the cross-section of Bi-DCB in the folding stable state. Galletly and Guest [20,21] developed the analytical model proposed by Iqbal and presented the beam model and the shell model. The predicted radius of the cross-section of the folding stable state is in good agreement with numerical simulation results, but there is a big difference with experimental results. Guest and Pellegrino [22] developed another inextensional bending model. In this analytical model, it is assumed that the deformation of the shell is uniform everywhere and the Gaussian curvature is always zero. Prediction results of this model are in good agreement with those of the above three analytical models [19-21]. The finite element method is an effective method with high prediction accuracy. Iqbal and Pellegrino [23] used the finite element method to study the folding stable state of the Bi-DCB, and gave the stress distribution and the radius of the cross-section of the folding stable state. Yang et al. [24] analysed the bistable deformation process of the Bi-DCB, and the results of finite element method and experiments are in good agreement.

According to previous reviews, the folding stable state of the Bi-DCB has obtained considerable attention. It is very important to fully understand the folding stable state of the Bi-DCB. However, the bistable mechanism of the Bi-DCB is extremely complex, such as the complex three-dimensional large deformation process, complex self-contact and complicated boundary conditions. How to determine the folding stable state of the Bi-DCB more accurately through experimental methods and numerical simulation has always been a very important issue. Therefore, in this paper, six Bi-DCB specimens were designed and manufactured, and the bistable deformation processes of the Bi-DCB specimens were achieved. Two nonlinear explicit finite element models (FEMs) were established to predict the folding stable state of the Bi-DCB. The folding stable state of the Bi-DCB (e.g., the geometric configuration of the cross-section and the failure index in the folding stable state) were determined and analyzed using experiments and two FEMs.

This paper is organised as follows: the functional mechanism of the folding stable state is introduced in Section 2; Bi-DCB specimens are prepared by the vacuum bag method and bistable experiments are carried out in Section 3; two FEMs to predict the folding stable state of the Bi-DCB are established in Section 4; the comparison between experiments and numerical simulations is discussed in Section 5; the key findings have been concluded in Section 6.

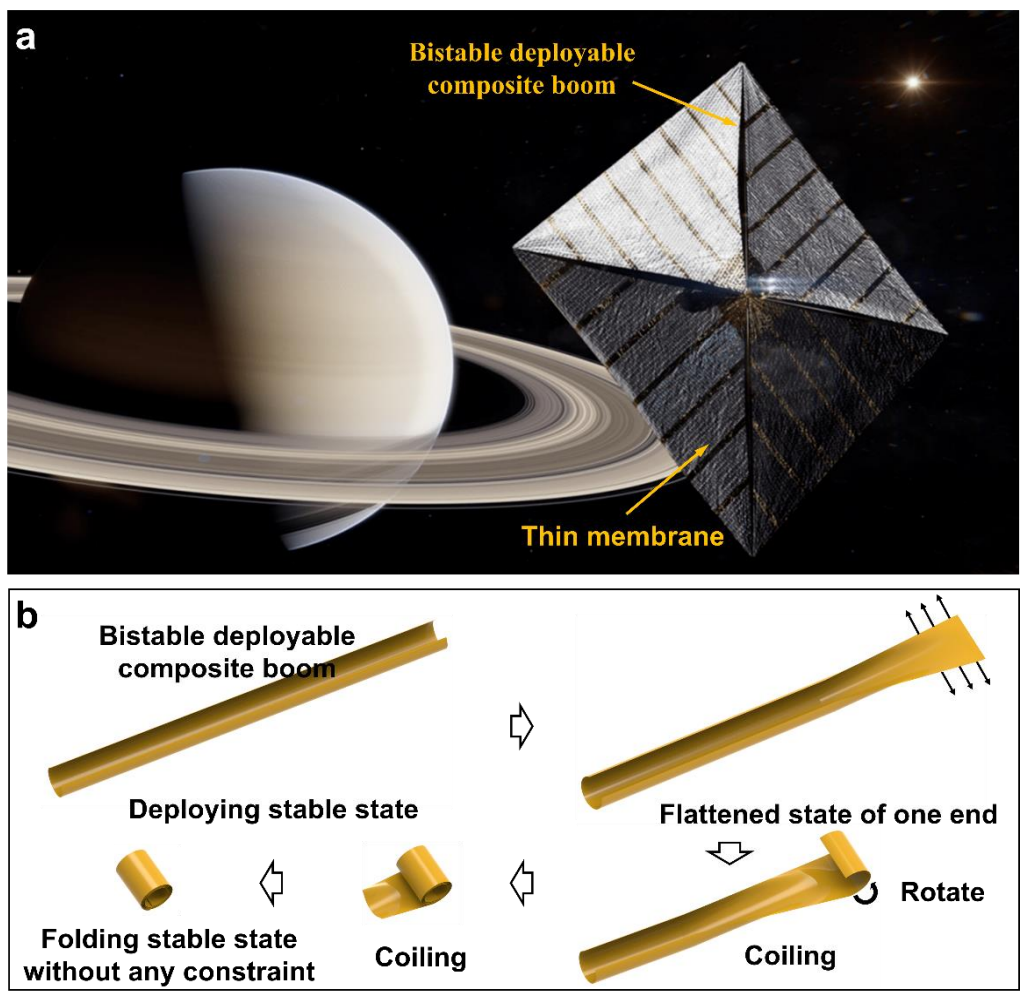

Fig. 1 Typical application of the Bi-DCB: (a) Deployable solar sail (b) Bistable deformation process.

## 2. Functional mechanism of bistable behavior

The deployable composite boom with symmetric layup can realize the folding and deploying functions. This deployable composite boom can be folded into a roll for packaging, and it also can recover to the initial configuration by releasing the stored elastic strain energy. However, when the deployable composite boom with symmetric layup is in the folding state, it has high strain energy and requires a restraint device to maintain the folding state. For the deployable composite boom with antisymmetric layup, the folding stable state and the deploying stable state can be achieved simultaneously through reasonable ply design, and the folding state can be maintained without external load constraints. This bistable behavior makes it possible to optimize the weight reduction of deployable devices. It has great application potential in the field of space deployable structures. The Bi-DCB mainly relies on the reasonable distribution of the bending stiffness to achieve the bistable function.

The curvature directions corresponding to the two stable states of the Bi-DCB are on the same side. A stable state is the initial deploying state, which has good bearing capacity; the other stable state is the folding state, where the Bi-DCB is closely and stably coiled together, as shown in Fig. 1b. Previous studies have shown that the main reason for the bistable behavior is Poisson effect [25]. When the Bi-DCB is stretched along the transverse direction, the transverse moment flattens the Bi-DCB. The large change of the transverse curvature causes the large longitudinal bending moment. Because $D_{16}$ and $D_{26}$ are zero in the bending stiffness matrix, there is no coupling effect between bending and twisting in the anti-symmetric laminate. To avoid the Bi-DCB ending up in a twisted configuration when it switches to the folding stable state, an anti-symmetric scheme is chosen to almost eliminate the coupling between bending and twisting, allowing for a compact rolling-up of the Bi-DCB. It should also be noted that the anti-symmetric laminate has a non-zero coupling stiffness matrix ***B***, but the resulting coupling effect between bending and stretching only has a weak influence on the bi-stability of the Bi-DCB [19].

## 3. Experiments

### 3.1 Materials

Bi-DCB specimens were prepared by ultra-thin T700/epoxy unidirectional reinforced prepreg. According to ASTM standards, T700/epoxy composite standard specimens were prepared and the basic mechanical properties were measured using MTS-100kN mechanical machine. Table 1 lists specifications and properties of T700/epoxy composite ply.

Table 1  Specifications and properties of T700/epoxy composite ply.

| Specifications and properties | Values |
|---|---|
| Longitudinal elastic modulus $E_1$ (GPa) | 128.62 |
| Transverse elastic modulus $E_2$ (GPa) | 7.52 |
| In-plane shear modulus $G_{12}$ (GPa) | 4.82 |
| Inter-laminar shear modulus $G_{13}$ (GPa) | 4.50 |
| Inter-laminar shear modulus $G_{23}$ (GPa) | 4.50 |
| Poisson's ratio $\nu_{12}$ | 0.314 |
| Longitudinal tensile strength $X_t$ (MPa) | 2103.44 |
| Transverse tensile strength $Y_t$ (MPa) | 75.97 |
| Longitudinal compressive strength $X_c$ (MPa) | 1233.65 |
| Transverse compressive strength $Y_c$ (MPa) | 181.46 |
| In-plane shear strength $S_{12}$ (MPa) | 216.36 |
| Density $\rho'$ (g/mm$^3$) | $1.60 \times 10^{-3}$ |
| Ply thickness (mm) | 0.03 |

**3.2 Specimen preparation**

The Bi-DCB can realize the conversion between the deploying stable state and the folding stable state by storing and releasing elastic strain energy. The Bi-DCB must have sufficient stiffness in the deploying stable state and sufficient flexibility in the folding stable state. The thickness is an important parameter that affects the stiffness and flexibility of the Bi-DCB. The thickness of the Bi-DCB needs to be in an appropriate range, and the large bending deformation in the bistable deformation process and the stiffness in the deploying state should be taken into account. In addition, the Bi-DCB should also be able to achieve multiple bistable functions, which requires that the Bi-DCB in the bistable deformation process at an appropriate stress level to ensure that the structure will not be damaged in the deformation process. Considering the fabrication process, cost and product quality, the vacuum bag method was selected to prepare Bi-DCB specimens.

According to the above design scheme and fabrication method, six types of Bi-DCB specimens were prepared. The detailed fabrication process of the Bi-DCB specimen is shown in Figs. 2a and 2b. The cure cycle included a heating ramp with a slope of 3.75°C per minute, followed by a curing phase that lasted 1 hour at a temperature of 150°C. Then, the specimens were gradually cooled down by cutting off the baking box. During the whole curing process, the pressure remained constant at 100kPa. Table 2 lists geometric parameters and stacking sequence of six types of Bi-DCB specimens. All specimens were cured using the same curing cycle. This study employed a slow heating method and delayed cooling process to alleviate residual thermal stresses and warping that may occur during the curing process of the Bi-DCBs with anti-symmetric laminates.

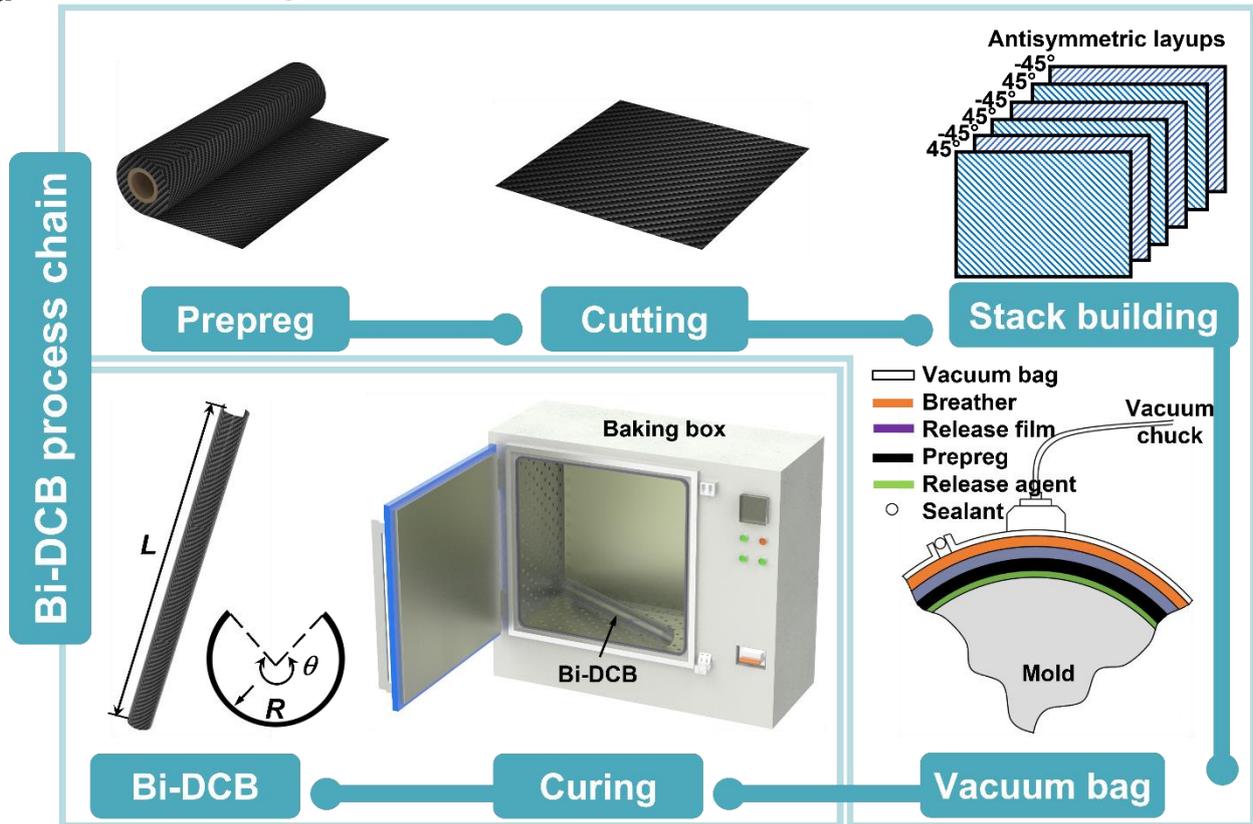

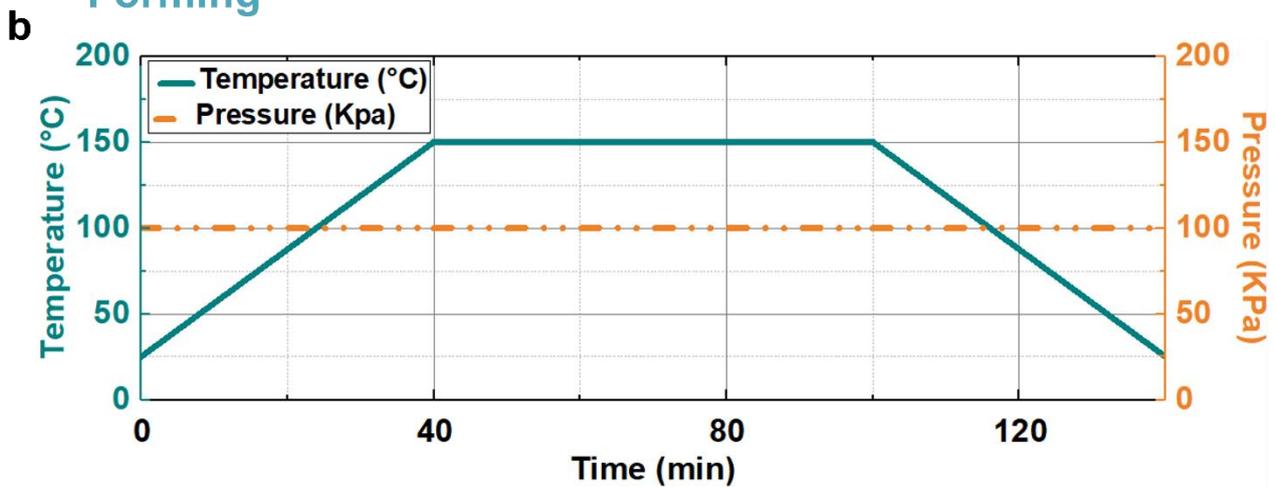

Fig. 2 Fabrication processes of Bi-DCB specimens: (a) Flow chart (b) Temperature and pressure during the curing process.

Table 2  Specifications of the six Bi-DCB specimens.

| No. | Radius $R$ (mm) | Central angle $\theta$ (deg) | Length $L$ (mm) | Stacking Configuration |
|---|---|---|---|---|
| 1 | 20 | 229 | 620 | [45/-45/45/-45/45/90/-45/45/-45/45/-45] |
| 2 | 20 | 229 | 400 | [45/-45/45/-45/0/45/-45/45/-45] |
| 3 | 20 | 229 | 200 | [45/-45/45/-45/90/45/-45/45/-45] |
| 4 | 27.5 | 167 | 650 | [45/-45/45/-45/0/90/45/-45/45/-45] |
| 5 | 27.5 | 167 | 650 | [45/-45/45/-45/45/-45/45/-45/45/-45] |
| 6 | 30 | 191 | 620 | [45/-45/45/-45/45/90/-45/45/-45/45/-45] |

**3.3 Bistable experiments**

Under the dry state at room temperature, the specimen was manually coiled to the folding stable state, as shown in Fig. 3a. To ensure that the Bi-DCB specimens did not suffer any damage during the bistable deformation process, three methods were employed to inspect the samples. Firstly, a sound inspection was conducted during the loading and unloading processes, monitoring for any signs of damage or failure sounds emanating from the specimens. Secondly, a visual inspection of the surface finish of the specimens was performed using a flashlight and magnifying glass. This involved carefully examining the surface for any signs of cracks or edge delamination, which could indicate structural damage. Thirdly, a tap inspection of the Bi-DCB specimens was conducted to check for any internal delamination issues. This involved lightly tapping the surface of the specimens with a small hammer and listening for any hollow sounds, which could indicate the presence of delamination. A camera was used to photograph the cross-section of the specimen No.1 in the folding stable state, and obtain the geometric configuration of the cross-section. The photos were imported into the GetData software to obtain the coordinate values of 30 data points on the neutral surface of the cross-section, $[\alpha_{e,i}, \rho_{e,i}]$ ($i=1,2,3...30$). Then, the coordinate values of the above data points were linearly fitted with Archimedes' helix, and the polar radii at the start-point and the end-point as well as the polar angle of Archimedes' helix were recorded.

Archimedes' helix can be described by the multinomial shape function in a polar coordinate system of $(\rho, \alpha)$, that is

$$\rho = a\alpha + b, \quad (\alpha \in [0, \alpha_1]) \tag{1}$$

where $a$ controls the distance between two adjacent circles and $b$ is the distance from the start-point to the origin of the polar coordinate system.

Using the least squares method to linear fit these data points $\left[\alpha_{e,i}, \rho_{e,i}\right](i=1,2,3...30)$, the values of $a$ and $b$ can be obtained as

$$\begin{cases} a = \dfrac{\sum\limits_{i=1}^{30}(\alpha_{e,i} - \overline{\alpha_e})(\rho_{e,i} - \overline{\rho_e})}{\sum\limits_{i=1}^{30}(\alpha_{e,i} - \overline{\alpha_e})^2} \\ b = \overline{\rho_e} - a\overline{\alpha_e} \end{cases} \quad (2)$$

where $\overline{\alpha_e}$ and $\overline{\rho_e}$ are the average values of these data points, and it is possible to show as

$$\begin{cases} \overline{\alpha_e} = (\sum\limits_{i=1}^{30} \alpha_{e,i})/30 \\ \overline{\rho_e} = (\sum\limits_{i=1}^{30} \rho_{e,i})/30 \end{cases} \quad (3)$$

According to geometric relationship, Eq. (1) should satisfy geometrical boundary condition as

$$\begin{cases} r_0 = b \\ r_1 = a\alpha_1 + b \end{cases} \quad (4)$$

where $r_0$ and $r_1$ are the polar radii at the start-point and the end-point of the Bi-DCB in the folding stable state, respectively.

The bistable experiments of six types of Bi-DCB specimens were conducted using the same method. The results show that the six types of Bi-DCB specimens can successfully achieve bistable function without damage. The deploying stable state and the folding stable state of six types of Bi-DCB specimens are shown in Fig. 3b. The detailed cross-section shapes of folding stable state are shown in Fig. 7 and Table 3.

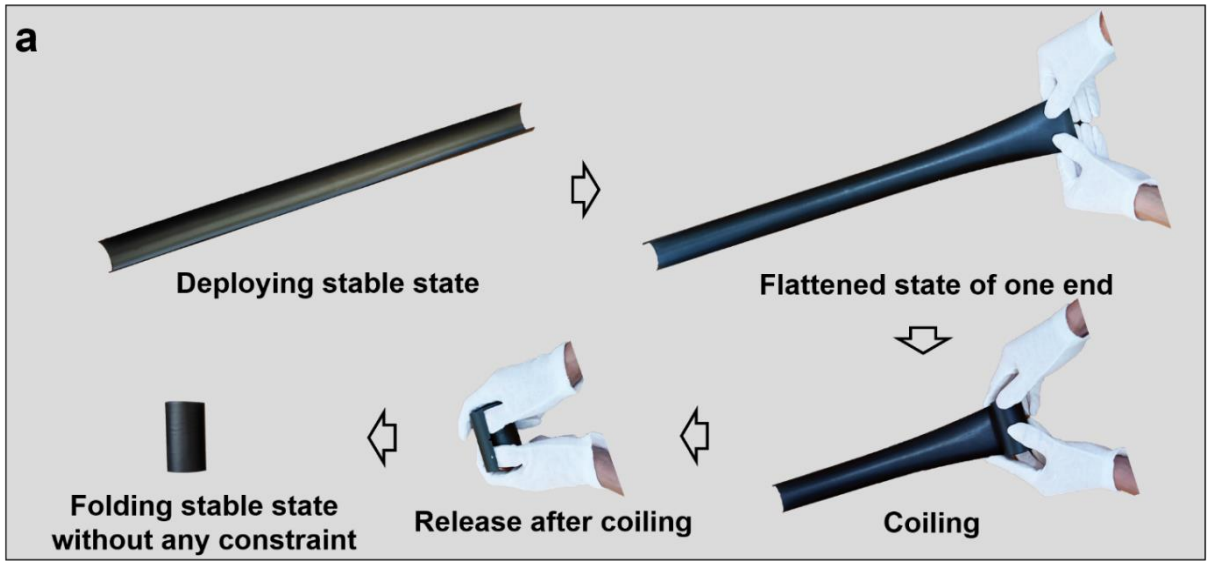

Fig. 3 Bistable experiments for the Bi-DCB: (a) Bistable deformation process of the Bi-DCB (b) Deploying and folding stable states of six Bi-DCB specimens.

## 4. Numerical simulation

In fact, in the bistable deformation process of the Bi-DCB, significant geometric changes and complex self-contact will lead to the singularities of the stiffness matrix, then resulting in the difficulty of solving convergence in dynamic analysis. The convergence problem can be solved by means of the explicit solver in the commercial finite element software ABAQUS [26]. Therefore, the geometrically nonlinear explicit dynamic analysis based on ABAQUS code were carried out. In this study, two FEMs were established to simulate the bistable deformation process of the Bi-DCB.

## 4.1 Finite element model 1

According to the geometrical configuration and specification of the Bi-DCB, a three-dimensional geometric nonlinear explicit FEM 1 of the Bi-DCB was established (shown in Fig. 4a). The 4-node doubly curved shell element with reduced integration (S4R) was utilized. Each node of this element has six degrees of freedom. The S4R element does not take into account inter-laminar shear stresses. In this study, the Bi-DCB has a very thin cross-section, and the in-plane stresses are dominant compared to the inter-laminar shear stresses. S4R element has been illustrated to be suitable for obtaining accurate simulation results in similar investigations [27,28]. In addition, this element is suitable for the analysis of nonlinear problems because it accounts for finite membrane strains and arbitrarily large rotation. The mesh size of the FEM can significantly affect the computing time and the accuracy of simulation results. Therefore, the mesh sensitivity analysis was performed to analyze the influence of mesh size on the folding stable state of the Bi-DCB. Considering the calculation accuracy and efficiency, the mesh sizes of the Bi-DCB and the roller were set to 5mm and 5mm, respectively (shown in Fig. 4b). The Bi-DCB was set as the material properties of T700/epoxy composite ply. Specifications and properties of T700/epoxy composite ply are listed in Table 1. The roller was set as the material properties of steel, in which the elastic modulus is 200GPa, Poisson's ratio is 0.3, and the density is $7.80 \times 10^{-3}$ g/mm$^3$. To improve the calculation efficiency, the roller was set as discrete rigid. In the FEM 1, the contact between the Bi-DCB and roller was set as the general contact. To make the Bi-DCB quickly reach the folding stable state, the friction coefficient of self-contact was set to 0.05. The bistable deformation process of the Bi-DCB was divided into three steps.

In the step 1, one end of the Bi-DCB was stretched until it was fully flattened, and the time of the step 1 was set to 0.5s. Displacements were applied on the two side edges AB and CD (shown in Fig. 4a), and the displacement directions were along the positive and negative directions of axis *y*, respectively. The applied displacement *l* can be determined using the initial geometry of the cross-section as shown in Fig. 4c, as follows:

$$l = \frac{R\theta}{2} - R\sin\frac{\theta}{2} \tag{5}$$

where $R$ and $\theta$ are the radius and central angle of the cross-section of the Bi-DCB, respectively.

The step 2 is used to simulate the coiling process of the Bi-DCB along the roller. A pressure of 1MPa was applied on the flattening region of the Bi-DCB, and then the flattening Bi-DCB was coiled by the roller. At the same time, an axial load of 2N is applied to the other end of the Bi-DCB. To generate lower kinetic energy in the coiling process, the time of the step 2 was set to 4s.

The step 3 is used to simulate the folding stable state of the fully coiled Bi-DCB. It is very important to select the roller diameter. If the roller diameter is too large or too small, which will make the Bi-DCB unable to reach the folding stable state in the step 3. Therefore, the closer the roller diameter is to diameter of the folding stable state, the easier it is for the Bi-DCB to reach the folding stable state. Iqbal et al. [19] established an analytical model for predicting the diameter of the folding stable state, which is in good agreement with experimental results. In the FEM 1, the roller diameter was set as the diameter of the folding stable state predicted by the above analytical model (shown in Fig. 4d). The diameter of the folding stable state can be derived as

$$d = \frac{2RD_{11}}{D_{12}} \tag{6}$$

where $D_{11}$ and $D_{22}$ are the bending stiffness of the laminate [29]. Using the above analytical model to determine the roller diameter was solely intended to expedite the attainment of the folding stable state. It should be emphasized that even if the roller diameter during the modelling process deviates by up to 10% from the predicted value by the analytical model, it will not affect the achievement of the folding stable state.

In the step 3, the constraints and all boundary conditions were removed to allow the Bi-DCB to reach the folding stable state. The time of the step 3 was set to 4s. The self-contact of the Bi-DCB was defined as general contact, in which the normal contact was set as hard contact and the tangential contact was set as smooth contact. The loading assignments of all three steps were set to table type. Internal energy, kinetic energy, Tsai-Hill failure index and maximum stress failure index were shown in Fig. 4b.

According to Fig. 4e, the internal energy (i.e., strain energy) in the Bi-DCB No.1 increases with the increase of tensile displacement and rotational displacement. In the step 3, the internal energy does not change, that is, the Bi-DCB No.1 reaches the folding stable state. Compared with internal energy, the kinetic energy obtained using the FEM 1 is very small and can be ignored. The bistable deformation process of the Bi-DCB No.1 simulated by the FEM 1 is shown in Fig. 4f. It is clear from Fig. 4f that, (i) the Bi-DCB No.1 can be folded stably along the roller, and finally reach the folding stable state; (ii) The Tsai-Hill failure indices of the Bi-DCB NO.1 in the bistable deformation process do not reach 1. Therefore, the Bi-DCB No.1 can realize bistable function without failure. Using the same method, the bistable deformation processes of the other five Bi-DCBs were simulated, and the analysis results were similar to those of Bi-DCB NO.1. Consistent with the experiments, the coordinate values of the data points of the numerical simulation results were linearly fitted with Archimedes' helix, and the polar radii at the start-point and the end-point as well as the polar angle of Archimedes' helix were recorded.

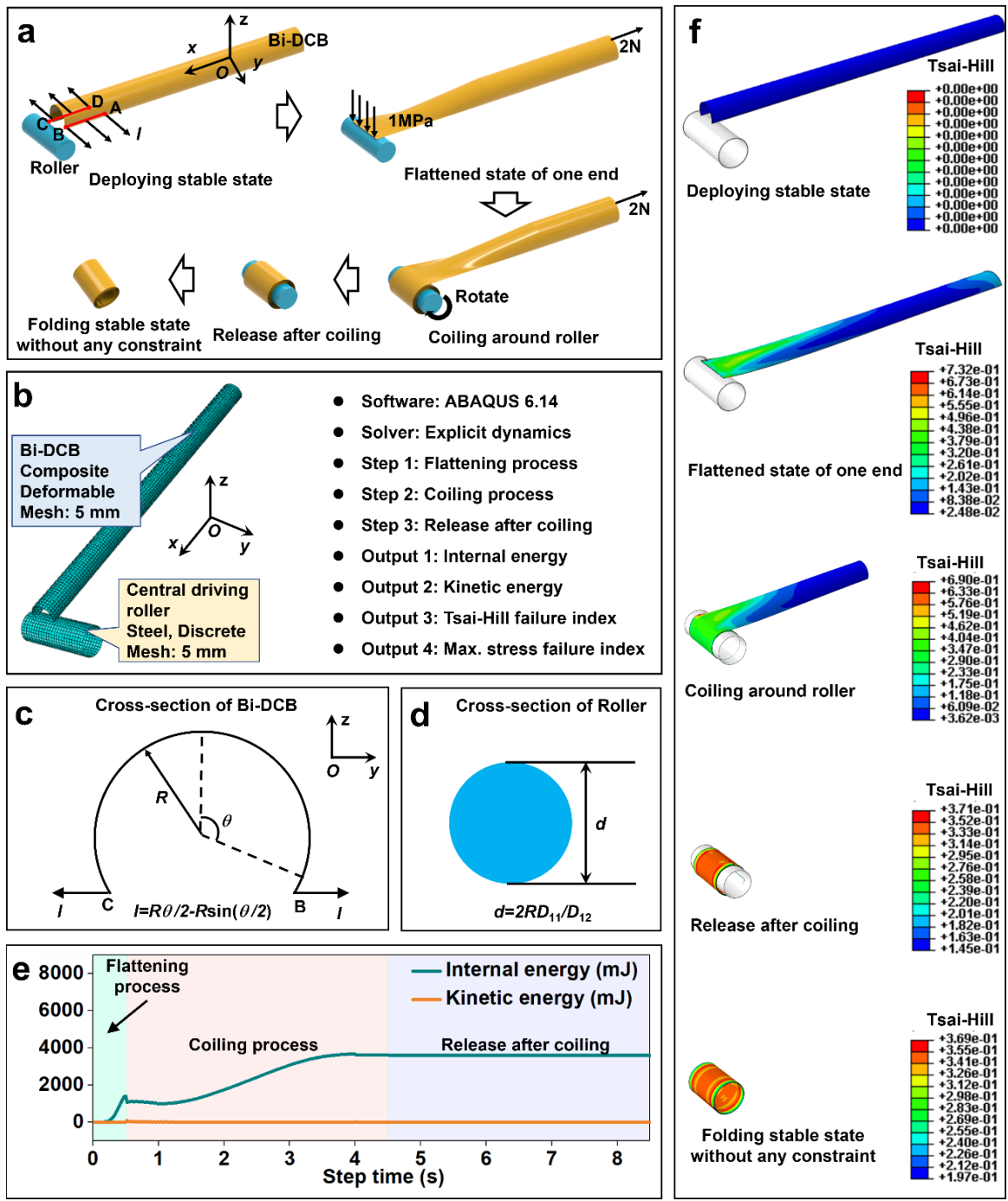

Fig. 4 FEM 1 for the Bi-DCB No.1: (a) Numerical simulation scheme (b) Mesh size (c) Applied displacement (d) Diameter of the roller (e) Energy changes in the three steps (f) Bistable deformation process through numerical simulation.

## 4.2 Finite element model 2

According to the geometrical configuration and specification of the Bi-DCB, a three-dimensional geometric nonlinear explicit FEM 2 of the Bi-DCB was established (shown in Fig. 5a). Similarly, S4R element was selected to model the Bi-DCB and rigid plate. The mesh sensitivity analysis was carried out for the elements with different sizes. Stable results can be obtained by setting the mesh sizes of the Bi-DCB and rigid plate to 5mm and 10mm, respectively (shown in Fig. 5b). Similar to the FEM 1, the Bi-DCB and rigid plate in the FEM 2 were set as the material properties of T700/epoxy composite ply and steel, respectively. In the FEM 2, the bistable deformation process of the Bi-DCB was divided into three steps.

In the step 1, one end of the Bi-DCB was stretched until it was fully flattened, and the time of the step 1 was set to 2s. Displacements were applied on the two side edges EF and GH (shown in Fig. 5a), and the displacement directions were along the positive and negative directions of axis $y$, respectively. Similar to the FEM 1, the applied displacement can be determined using the initial geometry of the cross-section of the Bi-DCB (shown in Fig. 5c).

The step 2 is used to simulate the coiling process of the Bi-DCB. To generate lower kinetic energy in the coiling process, the time of the step 2 was set to 15s. The contact between the Bi-DCB and the rigid plate was defined as general contact, in which the normal contact was set as hard contact and the tangential contact was set as smooth contact. The displacement loading method was adopted, and the rigid plate was used to assist the coiling of the Bi-DCB (shown in Fig. 5d). The displacement of the rigid plate in the $x$ direction can be expressed as

$$L^* = L - \frac{RD_{11}}{D_{12}} \tag{7}$$

where $L$ is the length of the Bi-DCB.

In the step 3, the constraints and all boundary conditions were removed to allow the Bi-DCB to reach the folding stable state. The time of the step 3 was set to 2s. The loading assignments of all three steps were set to table type. Internal energy, kinetic energy, Tsai-Hill failure index and maximum stress failure index were shown in Fig. 5b.

According to Fig. 5e, the internal energy (i.e., strain energy) in the Bi-DCB No.1 increases with the increase of tensile displacement and rotational displacement. In the step 3, the internal energy does not change, that is, the Bi-DCB No.1 reaches the folding stable state. Compared with internal energy, the kinetic energy predicted by the FEM 2 is very small and can be ignored. The bistable deformation process of the Bi-DCB No.1 simulated using the FEM 2 is shown in Fig. 5f. It is clear from Fig. 5f that, (i) the Bi-DCB No.1 can be folded stably with the movement of the rigid plate, and finally reach the folding stable state; (ii) The Tsai-Hill failure indices of the Bi-DCB NO.1 in the bistable deformation process do not reach 1. Therefore, the Bi-DCB No.1 can realize bistable function without failure. Using the same method, the bistable deformation processes of the other five Bi-DCBs were simulated, and the analysis results were similar to those of Bi-DCB NO.1. Consistent with the experiments, the coordinate values of the data points of the numerical simulation results were linearly fitted with Archimedes' helix, and the polar radii at the start-point and the end-point as well as the polar angle of Archimedes' helix were recorded.

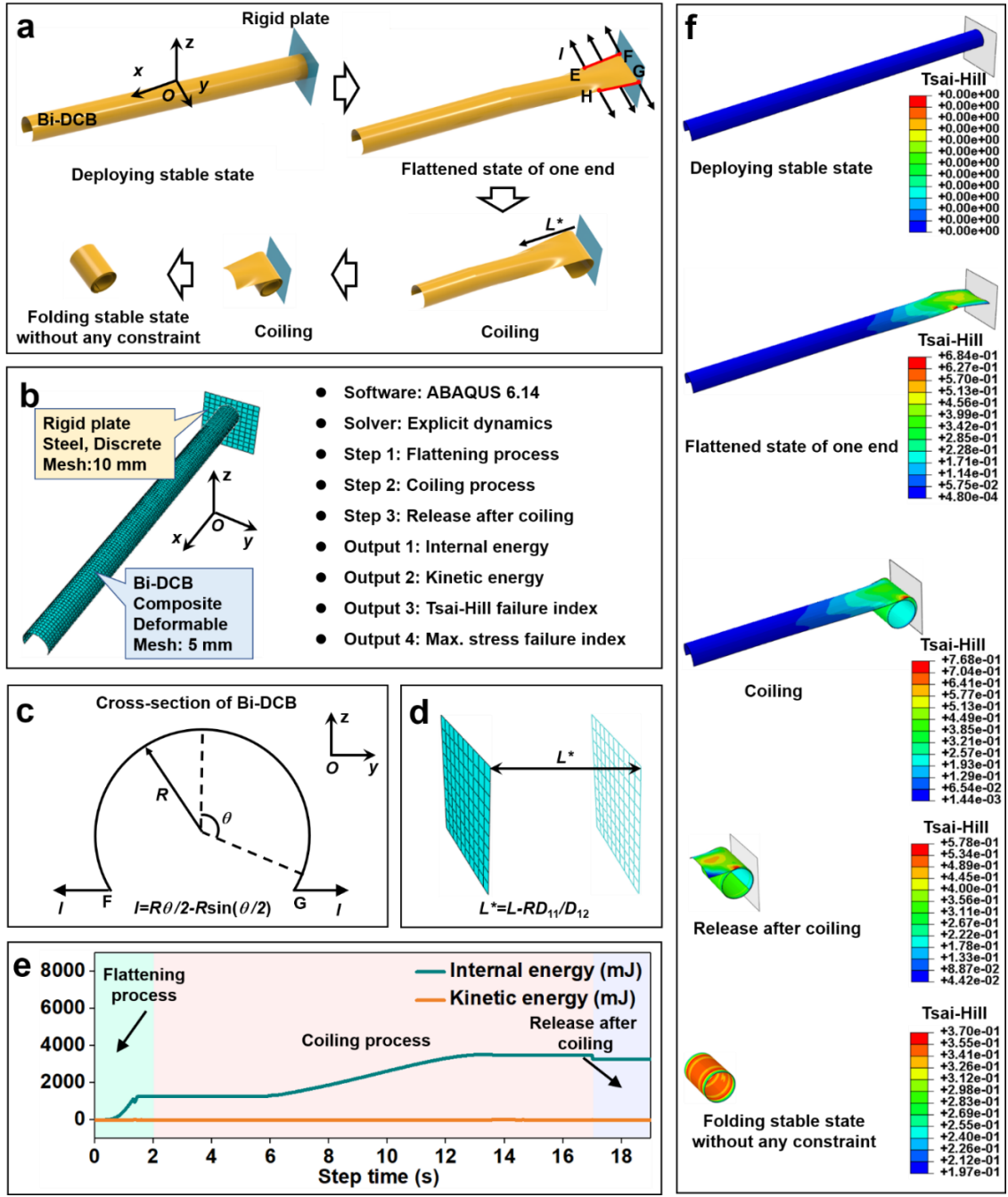

Fig. 5 FEM 2 for the Bi-DCB No.1: (a) Numerical simulation scheme (b) Mesh size (c) Applied displacement (d) Displacement of the rigid plate (e) Energy changes in the three steps (f) Bistable deformation process through numerical simulation.

## 4.3 Stress level

The failure index based on composite failure criterion can be employed to determine the stress level of the Bi-DCB in the folding stable state. When the failure index reaches or exceeds 1, the Bi-DCB fails; otherwise, it does not fail. Many failure criteria have been proposed to predict the failure of composites, such as the Tsai-Hill criterion and the maximum stress criterion. Many researchers applied the two failure criteria to investigate the failure of CFRP laminates [30-36]. Although the Tsai-Hill and maximum stress criteria cannot predict delamination issues, the interlaminar shear stress is very low and delamination does not occur in the bistable deformation process of the Bi-DCB. Therefore, the Tsai-Hill criterion and the maximum stress criterion were employed to predict the stress level of the Bi-DCB in the folding stable state.

The Tsai-Hill failure index $I_{f,1}$ is calculated by employing the following equations [29]:

$$I_{f,1}^2 = \frac{(\sigma_1^k)^2}{X_1^2} - \frac{\sigma_1^k \sigma_2^k}{X_2^2} + \frac{(\sigma_2^k)^2}{Y^2} + \frac{(\tau_{12}^k)^2}{S_{12}^2} \tag{8}$$

where

$$X_1 = \begin{cases} X_t & \text{if } \sigma_1^k > 0 \\ X_c & \text{if } \sigma_1^k < 0 \end{cases} \tag{9}$$

$$X_2 = \begin{cases} X_t & \text{if } \sigma_2^k > 0 \\ X_c & \text{if } \sigma_2^k < 0 \end{cases} \tag{10}$$

$$Y = \begin{cases} Y_t & \text{if } \sigma_2^k > 0 \\ Y_c & \text{if } \sigma_2^k < 0 \end{cases} \tag{11}$$

where $\sigma_1^k$ and $\sigma_2^k$ are the stresses of the $k^{th}$ ply in longitudinal and transverse direction respectively, $\tau_{12}^k$ is the shear stress of the $k^{th}$ ply in longitudinal and transverse direction, $X_t$ and $X_c$ are longitudinal tensile and compressive strength of composite ply, $Y_t$ and $Y_c$ are transverse tensile and compressive strength of composite ply, $S_{12}$ is the in-plane shear strength of composite ply, and $X_1$, $X_2$, and $Y$ are intermediate variables.

The maximum stress failure index $I_{f,2}$ is calculated by employing the following equations [29]:

$$I_{f,2} = \max\left\{\left|\frac{\sigma_1^k}{X_1}\right|, \left|\frac{\sigma_2^k}{Y}\right|, \frac{\tau_{12}^k}{S_{12}}\right\} \tag{12}$$

## 5. Results and discussion

Figs. 3a, 4f and 5f show the bistable deformation process of the Bi-DCB determined by experiments and two FEMs, and the three agree well. Fig. 6 illustrates the geometric configuration of the Bi-DCB in the folding stable state determined by experiments and two FEMs, and the three are in good agreement. Archimedes' helix was used to linearly fit the data points of the cross-section of the folding stable state determined by experiments and two FEMs. The fitted Archimedes' helix is shown in Fig. 7, and the polar radii at the start-point and the end-point as well as the polar angle at the end-point are listed in Table 3. According to Fig. 7 and Table 3, it is clear that the numerically predicted geometric configuration of the folding stable state, the polar radii at the start-point and the end-point as well as the polar angle at the end-point polar angle are in good agreement with experiment.

The two failure criteria in Section 4.3 (namely, Tsai-Hill criterion and maximum stress criterion) are applied to calculate the maximum failure index of each layer of six Bi-DCBs in the folding stable state, and the maximum failure indices are listed in Fig. 8 and Table 4. According to Fig. 8 and Table 4, among all the failure indices, the maximum values of FEM 1 and FEM 2 are 0.3694 and 0.3705 respectively. It means that the six Bi-DCBs can realize the bistable function without failure, which is consistent with experimental results. In conclusion, the two FEMs established in this study show good prediction accuracy for evaluating the folding stable state of the Bi-DCB, which proves the validity of the two FEMs.

For the six Bi-DCBs, the polar radii at the start-point and the end-point of Archimedes' helix predicted by the two FEMs are higher than experimental results, while the polar angle at the end-point polar angle predicted by the FEMs are lower than experimental results. There is a possible reason for the difference between prediction results of the two FEMs and experimental results. In the manufacturing process of the Bi-DCB specimens, residual thermal stresses would be inevitably generated, which results in an evident discrepancy for the deformation of the laminates. In future research, a comprehensive consideration of materials, manufacturing processes, and structural design can be made to adopt appropriate techniques and methods to reduce the impact of residual thermal stress, thereby improving the reliability and mechanical property of the Bi-DCB. Controlling the heating and cooling rates, optimizing the layering sequence and thickness, and optimizing the geometric shape are effective methods for reducing the residual thermal stress in asymmetrically laminated Bi-DCB during the curing process. First, controlling the heating and cooling rates is crucial. In the manufacturing process, slower heating and cooling rates can be used to allow the material to gradually adapt to temperature changes. Second, optimizing the layering sequence and thickness can also reduce residual thermal stress. Adjusting the thickness and layering sequence of different plies can reduce the production of residual thermal stress. Finally, optimizing the geometric shape can also reduce the impact of residual thermal stress. The ply angle can be optimized to reduce stress concentration and the production of residual thermal stress. For example, when designing the geometric shape of a Bi-DCB, a structure design that reduces internal stress can be considered to reduce the production of residual thermal stress.

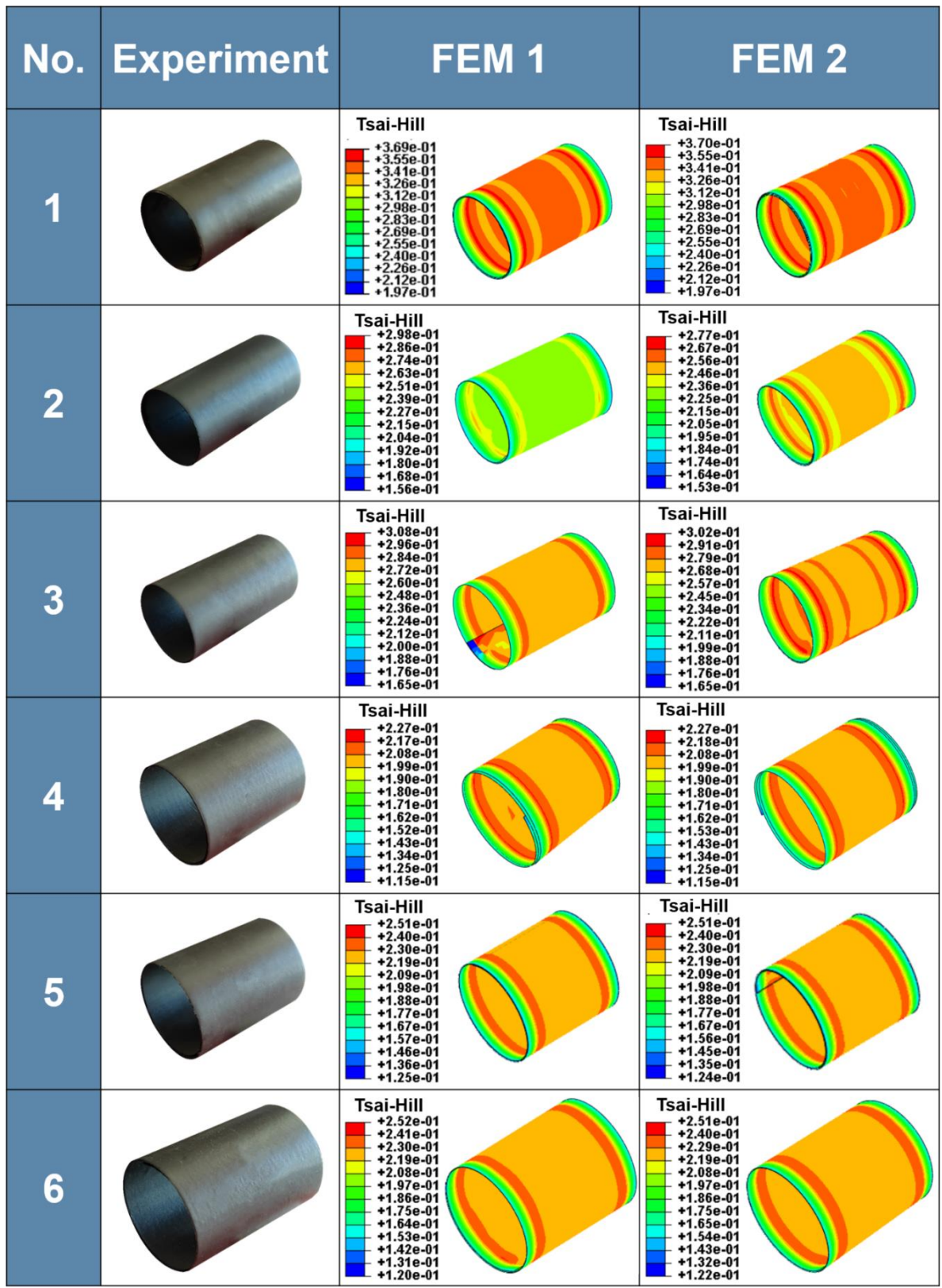

Fig. 6  Geometric configuration of six Bi-DCBs in the folding stable state.

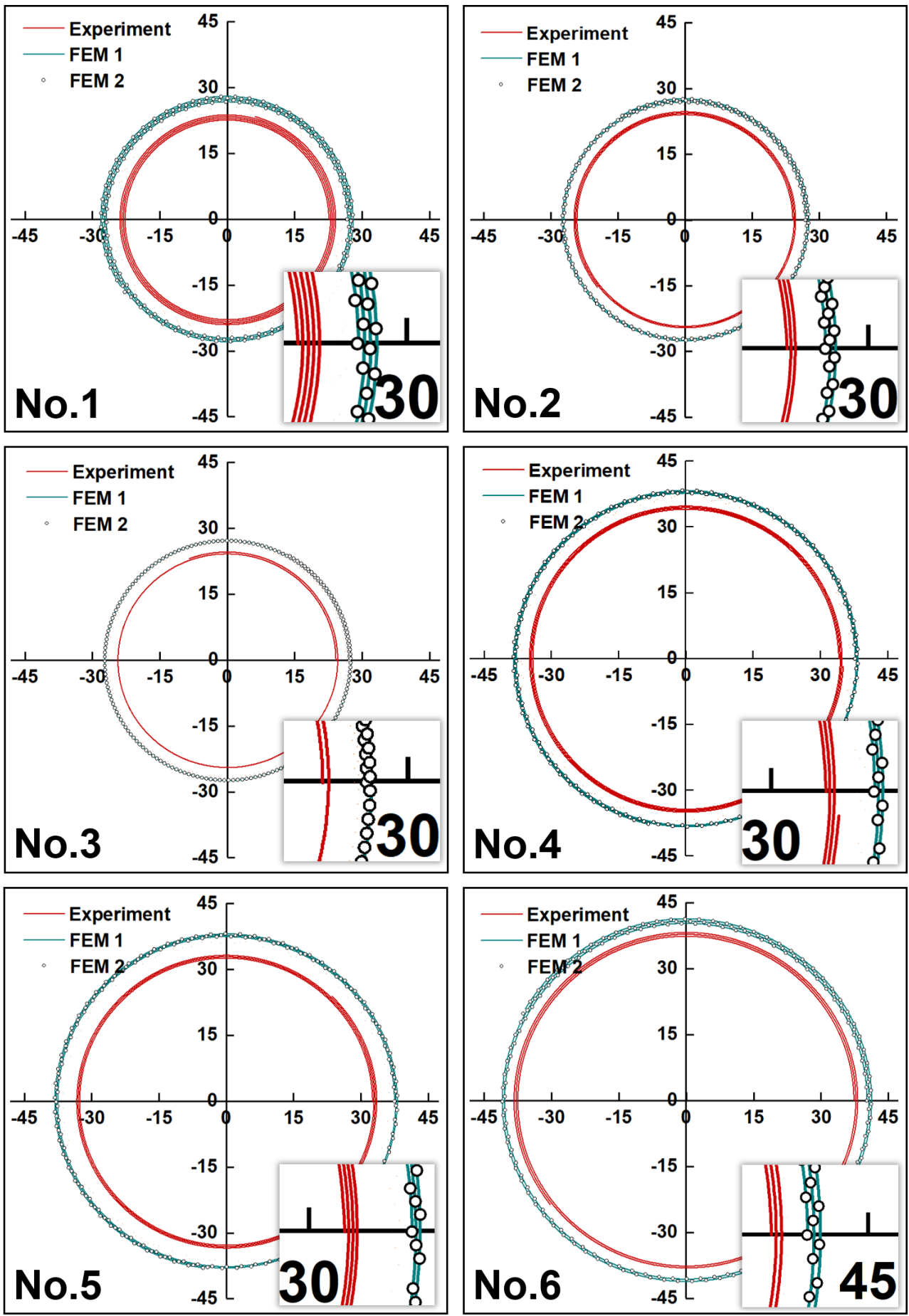

Fig. 7 Geometric configuration of the cross-section of six Bi-DCBs in the folding stable state.

Table 3  Polar radii at the start-point and the end-point as well as polar angle of six Bi-DCBs in the folding stable state.

| Geometrical parameter | | | No.1 | No.2 | No.3 | No.4 | No.5 | No.6 |
|---|---|---|---|---|---|---|---|---|
| $r_0$ (mm) | Experiment | | 22.63 | 24.04 | 24.13 | 34.08 | 32.59 | 37.31 |
| | FEM 1 | Value | 26.75 | 26.89 | 27.16 | 37.75 | 37.49 | 40.21 |
| | | Error | 18.21% | 11.86% | 12.56% | 10.77% | 15.04% | 7.77% |
| | FEM 2 | Value | 26.66 | 26.85 | 27.15 | 37.68 | 37.47 | 40.15 |
| | | Error | 17.81% | 11.69% | 12.52% | 10.56% | 14.97% | 7.61% |
| $r_1$ (mm) | Experiment | | 24.23 | 24.87 | 24.66 | 35.08 | 33.58 | 38.41 |
| | FEM 1 | Value | 28.23 | 27.75 | 27.52 | 38.57 | 38.32 | 41.47 |
| | | Error | 16.51% | 11.58% | 11.60% | 9.95% | 14.11% | 7.97% |
| | FEM 2 | Value | 28.18 | 27.67 | 27.44 | 38.55 | 38.31 | 41.35 |
| | | Error | 16.30% | 11.26% | 11.27% | 9.89% | 14.09% | 7.65% |
| $\alpha_1$ (rad) | Experiment | | 26.46 | 16.36 | 8.20 | 18.80 | 19.65 | 16.38 |
| | FEM 1 | Value | 22.55 | 14.64 | 7.32 | 17.03 | 17.15 | 15.07 |
| | | Error | 14.78% | 10.51% | 10.73% | 9.41% | 12.72% | 8.00% |
| | FEM 2 | Value | 22.61 | 14.67 | 7.33 | 17.05 | 17.15 | 15.11 |
| | | Error | 14.55% | 10.33% | 10.61% | 9.31% | 12.72% | 7.75% |

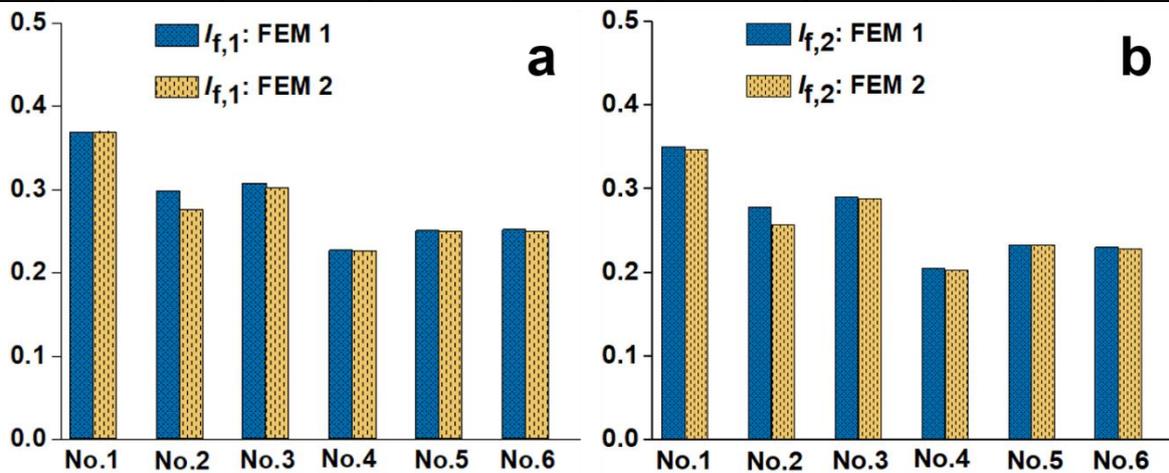

Fig. 8  Comparison of the Failure indices of six DCBs in the folding stable state: (a) Tsai-Hill failure index (b) Maximum stress failure index.

Table 4  Failure indices of six DCBs in the folding stable state.

| Failure index | | No.1 | No.2 | No.3 | No.4 | No.5 | No.6 |
|---|---|---|---|---|---|---|---|
| $I_{f,1}$ | FEM 1 | 0.3694 | 0.2981 | 0.3079 | 0.2266 | 0.2505 | 0.2520 |
| | FEM 2 | 0.3705 | 0.2769 | 0.3024 | 0.2270 | 0.2509 | 0.2509 |
| $I_{f,2}$ | FEM 1 | 0.3514 | 0.2786 | 0.2907 | 0.2058 | 0.2333 | 0.2304 |
| | FEM 2 | 0.3477 | 0.2577 | 0.2880 | 0.2034 | 0.2328 | 0.2284 |

## 6. Conclusions

This paper aimed to study the folding stable state of the Bi-DCB by combining experimental and numerical simulation methods. Three important results emerging from the research are as follows:

(1) The bistable experiments were carried out on the Bi-DCB specimens, and the geometric configuration of the folding stable state was determined using Archimedes' helix.

(2) Two FEMs were established for predicting the folding stable state of the Bi-DCB, and numerical results of the two FEMs are in good agreement with experimental results, including the bistable deformation process and the folding stable state.

(3) Based on Tsai-Hill criterion and maximum stress criterion, the predicted maximum failure indices of the FEMs are 0.3694 and 0.3705, respectively. It is indicated that the six Bi-DCBs can realize the bistable function without failure, which is consistent with experiments.

As mentioned in the introduction, the folding stable state of the Bi-DCB is complex, such as the complex three-dimensional deformation process, contact and boundary conditions. In this paper, experimental and numerical simulation methods provide a solution to effectively solve the above problems, which is the main innovation of this paper. Compared with the complex experimental and numerical simulation methods, the analytical modeling is simple and efficient. Further study is necessary to study the folding stable state of the Bi-DCB through analytical modeling. Two reasons are given to explain the difference between prediction results of the FEMs and experimental results. Quantitatively research on the above two reasons will be performed in the future.

## Credit authorship contribution statement

Tian-Wei Liu: Investigation, Conceptualization, Data Curation, Methodology, Software, Formal analysis, Validation, Writing-original draft. Jiang-Bo Bai: Conceptualization, Supervision, Resources, Funding acquisition, Project administration, Methodology, Writing-review & editing. Hao-Tian Xi:

Software, Validation. Nicholas Fantuzzi: Supervision, Writing-review & editing. Guang-Yu Bu: Validation. Yan Shi: Supervision, Writing-review & editing.

## Acknowledgements

This project was supported by the National Natural Science Foundation of China (Grant No. 52275231 and Grant No. 51875026) and the National Defence Basic Scientific Research Program of China (Grant No. JCKY2019205C002). The first author was supported by the Academic Excellence Foundation of BUAA for PhD Students and the China Scholarship Council (Grant No. 202106020152).

## References

[1] Foster CL, Tinker ML, Nurre GS, et al. Solar-array-induced disturbance of the Hubble space telescope pointing system. Journal of Spacecraft and Rockets, 1995, 32(4):634-644.

[2] Sawada H, Mori O, Okuizumi N, et al. Mission report on the solar power sail deployment demonstration of IKAROS. 52nd AIAA/ASME/ASCE/AHS/ASC Structures, Structural Dynamics and Materials Conference 19th AIAA/ASME/AHS Adaptive Structures Conference 13t, 2011.

[3] Delleur A, Kerslake.T. Managing ISS US Solar Array Electrical Hazards for SSU Replacement via EVA. 2nd International Energy Conversion Engineering Conference., 2004.

[4] Liu TW, Bai JB, Fantuzzi N. Folding behaviour of the thin-walled lenticular deployable composite boom: Analytical analysis and many-objective optimization. Mechanics of Advanced Materials and Structures, 2022:2053766.

[5] Sharma G, Omkar SN, Kumar H, et al. Design, modeling, analysis and development of deployable tubular metallic booms for space application. Materials Today: Proceedings, 2020, 28:2463-2470.

[6] Mar J, Garrett T. Mechanical design and dynamics of the Alouette spacecraft. Proceedings of the IEEE, 2005, 57(6):882-896.

[7] Bai JB, Liu TW, Yang GH, et al. A variable camber wing concept based on corrugated flexible composite skin. Aerospace Science and Technology, 2023: 108318.

[8] Guo SX, Li XQ, Liu TW, et al. Parametric study on low-velocity impact (LVI) damage and compression after impact (CAI) strength of composite laminates. Polymers, 2022, 14:5200.

[9] Liu TW, Bai JB. Folding behaviour of a deployable composite cabin for space habitats - Part 1: Experimental and numerical investigation. Composite Structures, 2022, 302:116244.

[10] Liu TW, Bai JB. Folding behaviour of a deployable composite cabin for space habitats - Part 2: Analytical investigation. Composite Structures, 2022, 297:115929.

[11] Liu TW, Bai JB, Fantuzzi N, et al. Multi-objective optimisation designs for thin-walled deployable composite hinges using surrogate models and Genetic Algorithms. Composites Structures, 2022, 280:114757.

[12] Bai JB, Xiong JJ, Gao JP, et al. Analytical solutions for predicting in-plane strain and interlaminar shear stress of ultra-thin-walled lenticular collapsible composite tube in fold deformation. Composite Structures, 2013, 97:64-75.

[13] Liu TW, Bai JB, Lin QH, et al. An analytical model for predicting compressive behaviour of composite helical Structures: Considering geometric nonlinearity effect. Composites Structures, 2021, 255:112908.

[14] Bai JB, Liu TW, Wang ZZ, et al. Determining the best practice – Optimal designs of composite helical structures using Genetic Algorithms. Composites Structures, 2021, 268:113982.

[15] Laurenzi S, Rufo D, Sabatini M, et al. Characterization of deployable ultrathin composite boom for microsatellites excited by attitude maneuvers. Composite Structures, 2019, 220:502-509.

[16] Laurenzi S, Rufo D, Sabatini M, et al. Characterization of deployable ultrathin composite boom for microsatellites excited by attitude maneuvers. Composite Structures, 2019, 220:502-509.

[17] Yang H, Guo H, Liu R, et al. Coiling and deploying dynamic optimization of a C-cross section thin-walled composite deployable boom. Structural and Multidisciplinary Optimization, 2020, 61(4):1731-1738.

[18] Fernandez JM，Visagie L，Schenk M，et al. Design and development of a gossamer sail system for deorbiting in low earth orbit. Acta Astronautica, 2014, 103:204-225.

[19] Iqbal K, Pellegrino S, Daton-Lovett A. Bistable composite slit tubes. IUTAM-IASS Symposium on Deployable Structures. Cambridge, UK, 1998.

[20] Galletly DA, Guest SD. Bistable composite slit tubes. I. A beam model. International Journal of Solids and Structures, 2004, 41(16-17):4517-4533.

[21] Galletly DA, Guest SD. Bistable composite slit tubes. II. A shell model. International Journal of Solids and Structures, 2004, 41(16-17): 4503-4516.

[22] Guest SD, Pellegrino S. Analytical models for bistable cylindrical shells. Proceedings of the Royal Society A: Mathematical, Physical and Engineering Sciences, 2006:839-854.

[23] Iqbal K, Pellegrino S. Bi-stable composite shells. 41st Structures, Structural Dynamics, and Materials Conference and Exhibit, 2000.

[24] Yang LY, Tan HF, Cao ZS. Modeling and analysis of the ploy region of bistable composite cylindrical shells. Composite Structures, 2018, 192:347-354.

[25] Daton-Lovett AJ, Compton-Bishop QM, Curry RG. Deployable structures using bi-stable reeled composites. Conference on active materials: Behavior and mechanics, 2000.

[26] Abaqus, 2014 Analysis User Maual Version 6.14 Dassault System.

[27] Bai JB, Chen D, Xiong JJ, et al. Folding analysis for thin-walled deployable composite boom. Acta Astronautica. 2019, 159:622-636.

[28] Yang H, Liu L, Guo H, et al. Wrapping dynamic analysis and optimization of deployable composite triangular rollable and collapsible booms. Structural and Multidisciplinary Optimization, 2019, 59: 1371-1383.

[29] Shen GL, Hu GK. Mechanics of composite materials. Tsinghua University Press, Beijing, 2006.

[30] Fernandes P, Sousa B, Marques R, et al. Influence of relaxation on the deployment behaviour of a CFRP composite elastic-hinge. Composite Structures, 2020:113217.

[31] Fragassa C, Pavlovic A, Minak G. On the structural behaviour of a CFRP safety cage in a solar powered electric vehicle. Composite Structures, 2020, 252:112698.

[32] Golewski P, Sadowski T. Description of thermal protection against heat transfer of carbon fiber reinforced plastics (CFRP) coated by stiffened ceramic mat (TBC). Composite Structures, 229:111489.

[33] Mohee FM, Al-Mayah A. Development of an innovative prestressing CFRP plate anchor: numerical modelling and parametric study. Composite Structures, 2017, 177:1-12.

[34] Mohee FM, Al-Mayah A. Effect of modulus of elasticity and thickness of the CFRP plate on the performance of a novel anchor for structural retrofitting and rehabilitation applications. Engineering Structures, 2017, 153:302-316.

[35] Liu TW, Bai JB, Li SL, et al. Large deformation and failure analysis of the corrugated flexible composite skin for morphing wing. Engineering Structures, 2023, 278: 115463.

[36] Liu T W, Bai J B, Fantuzzi N. Analytical models for predicting folding behaviour of thin-walled tubular deployable composite boom for space applications. Acta Astronautica, 2023, 208:167-178.